\documentclass{iopart}

\usepackage{graphicx}
\usepackage{harvard}

\def\jpb{J. Phys. B }

\newcommand{\Sec}[1]{Section~\ref{#1}}
\newcommand{\Fig}[1]{Fig.~\ref{#1}}

\newcommand{\Liplus}{Li$^{+}$}
\newcommand{\singS}{$1s2s\ ^1\!S$}
\newcommand{\tripS}{$1s2s\ ^3\!S$}

\begin{document}

\title{Electron-impact ionization of the metastable excited states of Li$^{+}$}

\author{J C Berengut$^1$, S D Loch$^1$, C P Ballance$^2$ and M S Pindzola$^1$}
\address{$^1$Department of Physics, Auburn University, Auburn, AL 36849, USA}
\address{$^2$Department of Physics, Rollins College, Winter Park, Florida 32789, USA}

\date{\today}

\begin{abstract}

Electron-impact ionization cross sections for the \singS\ and \tripS\ metastable states of \Liplus\ are calculated using both perturbative distorted-wave and non-perturbative close-coupling methods. Term-resolved distorted-wave calculations are found to be approximately 15\% above term-resolved \mbox{$R$-matrix} with pseudostates calculations. On the other hand, configuration-average time-dependent close-coupling calculations are found to be in excellent agreement with the configuration-average \mbox{$R$-matrix} with pseudostates calculations. The non-perturbative \mbox{$R$-matrix} and close-coupling calculations provide a benchmark for experimental studies of electron-impact ionization of metastable states along the He isoelectronic sequence.

\end{abstract}

\submitto{\jpb}

\maketitle

\section{\label{sec:intro} Introduction}

Electron-impact ionization of low-charged ions is an important component of many processes in astrophysics and controlled fusion plasmas. This topic has therefore generated considerable interest, both experimental and theoretical. Previous studies of ground-state He \cite{pindzola98pra,pindzola04pra}, as well as He-like Li \cite{pindzola00praA} and Be \cite{colgan03pra}, have shown that both the $R$-matrix with pseudo-states (RMPS) method \cite{bartschat98cpc} and time-dependent close-coupling (TDCC) methods \cite{pindzola96praA,pindzola96praB} give cross-sections that are in very good agreement with experiment.

Far less work has been done on the electron-impact ionization cross-section of metastable ions. One important example is excited He (\tripS), that has been studied both in experiment and theory. The experiment, performed by \citeasnoun{dixon76jpb}, is in reasonable agreement at high energies with Bethe-Born calculations that pre-date the experiment \cite{briggs71pra}, and is in good agreement with first-order plane-wave Born results generated shortly afterwards \cite{tonthat77jpb}. More recently, however, several non-perturbative calculations have shown a large discrepancy with experiment \cite{colgan02pra,bartschat02jpb,fursa03jpb}. These papers used TDCC, convergent close-coupling, and $R$-matrix with pseudostates (RMPS). All of these calculations are in agreement, and all differ from the Dixon experiment by a factor of two, even at high energies. The helium discrepancy motivates us to see whether the situation improves with metastable He-like Li.

Further motivation comes from plasma modeling. It has been previously shown that ionization from the metastable \Liplus\ can dominate the effective ionization rate coefficient even at relatively low electron densities ($10^{10}$~cm$^{-3}$) \cite{loch04pre}. However, all existing data on the excited states of \Liplus\ is based on distorted-wave calculations, which can get progressively worse as the term energy gets higher; this was demonstrated in H-like ions, including Li$^{2+}$, see \citeasnoun{griffin05jpb}. Thus it is important to test the accuracy of the ionization data from excited states and to determine the effect that such data has on plasma modeling.

In this paper we build upon a previous study of the \Liplus\ ground state \cite{pindzola00praA} to examine the electron-impact ionization cross-section of metastable \singS\  and \tripS~\Liplus. We compare the perturbative distorted-wave method with the non-perturbative RMPS and TDCC methods. The cross-sections calculated using the non-perturbative methods converge to the same value in \Liplus, while the distorted-wave method gives results that are significantly higher. This study provides a benchmark for future experimental studies of electron-impact ionization from metastable states along the He isoelectronic sequence. In the following section we discuss the various methods we have used to calculate electron-impact ionization cross-sections of metastable \Liplus. The results of these calculations are presented and compared in \Sec{sec:results}.

\section{\label{sec:theory} Theory}

\subsection{\label{ssec:dw} Distorted-wave method}

The distorted-wave (DW) cross sections are calculated from a triple partial-wave expansion of the first-order perturbation theory scattering amplitude. Both direct and exchange components of the scattering amplitude are included. The incoming and outgoing electrons are calculated in a $V^N$ potential, while the ejected electron is calculated in a $V^{N-1}$ potential, where $N=2$ is the number of electrons in the target ion \cite{younger85book}. The DW potential for all continuum electrons is made up of a configuration-average Hartree potential for the direct interaction, and a local semiclassical approximation for the exchange interaction.

For the term-resolved distorted-wave calculations, the bound orbitals are calculated using Fischer's multi-configuration Hartree-Fock code \cite{fischer97book}. A double configuration ($1s^2$ and $1s2s$) calculation for the \singS\ term yields an energy of -5.036 a.u. and an ionization potential of 14.59 eV, in good agreement with the experimental value of 14.86 eV \cite{NIST}. A single configuration ($1s2s$) calculation for the \tripS\ term yields an energy of -5.109 a.u. and an ionization potential of 16.58 eV, in very good agreement with the experimental value of 16.60 \cite{NIST}.

For the configuration-average DW calculations, the bound orbitals are calculated using \possessivecite{cowan81book} Hartree-Fock code. Calculation for the $1s2s$ configuration yields an energy of 16.07 eV, in good agreement with the experimental value of 16.16 eV.


\subsection{\label{ssec:rmps} $R$-matrix with pseudostates method}

The codes used in this work are based on the serial codes published by \citeasnoun{berrington95cpc}, with modifications for the addition of pseudostates \citeaffixed{gorczyca97jpb}{see, e.g.}, and parallelization described in \citeasnoun{mitnik99jpb}, \citeasnoun{mitnik03jpb}, and \citeasnoun{ballance04jpb}. In this method, the high Rydberg states and the target continuum are represented by a set of orthogonalized Laguerre radial wavefunctions: the so-called pseudostates. The ionization cross-sections are determined by summing over excitations above the ionization threshold, including all single-electron excitations to the pseudostates as well as doubly excited states. 

Our RMPS basis used spectroscopic orbitals up to $n = 3$ and pseudostates from $n = 4$ to $n=14$, with a maximum angular momentum of $l=4$. The pseudostates with $l=4$ and $11 \leq n \leq 14$ were omitted to keep the size of the calculation manageable. For the incoming electron, partial waves from $l = 0$ to $13$ were calculated using the $R$-matrix method with exchange included. This was topped up from $l=14$ with methods described by \citeasnoun{burgess70ptrsa} \citeaffixed{burgess74jpb}{see also}.

\subsection{\label{ssec:tdcc} Time-dependent close-coupling method}

We use the ``frozen core'' approximation that was previously used in electron-ionization from metastable helium, where we freeze the $1s$ electron (see \citeasnoun{colgan02pra} and references therein). The frozen $1s$ orbital is the ground state of the hydrogenic Li$^{2+}$ ion. A complete set of orbitals are then obtained by diagonalization of the single-particle Hamiltonian, including the direct term of the Hartree potential, and a local approximation to the exchange interaction. The local exchange potential was scaled to give single-particle energies close to experiment.

The initial two-electron wavefunction, $P^{LS}_{l_1 l_2} (r_1, r_2, t=0)$, is an antisymmetrized product of a radial wavepacket and the target $2s$ radial orbital, with a particular $LS$ symmetry. The propagation in time is governed by the Schr\"odinger equation, which we write as a set of time-dependent close-coupled partial differential equations
\begin{eqnarray}
i\frac{\partial P^{LS}_{l_1 l_2} (r_1, r_2, t)}{\partial t}
 &= & T_{l_1 l_2} (r_1, r_2) P^{LS}_{l_1 l_2} (r_1, r_2, t) \nonumber \\
 &   & + \sum_{l_1', l'_2} U^L_{l_1 l_2, l'_1 l'_2} (r_1, r_2) P^{LS}_{l'_1 l'_2} (r_1, r_2, t) \ ,
\end{eqnarray}
where $T_{l_1 l_2} (r_1, r_2)$ is the single particle Hamiltonian, which includes kinetic energy, nuclear, direct and local exchange operators, while $U^L_{l_1 l_2, l'_1 l'_2} (r_1, r_2)$ couples the $(l_1 l_2)$ scattering channels. Some time after the collision, the two-electron radial wavefunctions are projected onto products of the single-particle orbitals to calculate the probability of excitation. The ionization probabilities, and hence the ionization cross-sections, are found by subtracting the probability of any electrons being bound from unity.

\section{\label{sec:results} Results}

\begin{figure}[tb]
\centering
\includegraphics[width=0.75\textwidth]{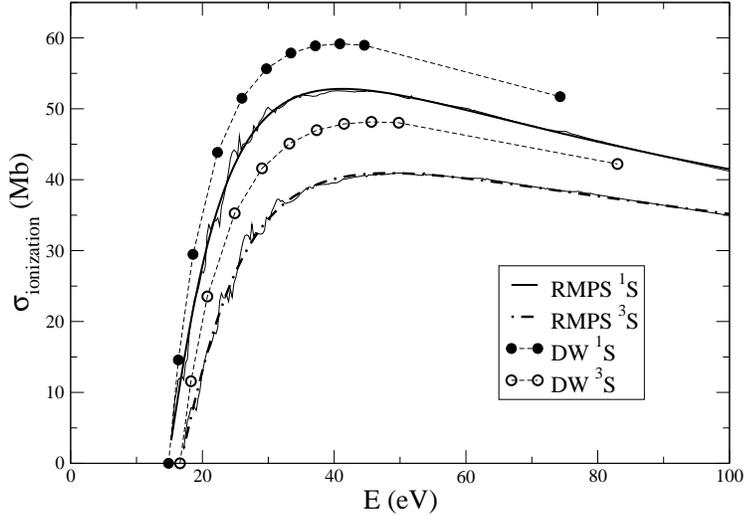}
\caption{\label{fig:terms} Term-resolved \singS\ and \tripS\ electron-impact ionization cross-sections, $\sigma_\textrm{ionization}$, against incident electron energy, $E$. We show the raw RMPS output (light solid lines) along with fits from \Eref{eq:fits} (heavy lines). The raw distorted-wave data is also shown (circles). \mbox{($1\,\textrm{Mb} = 10^{-18}\,\textrm{cm}^{2}$)}}
\end{figure}

The term resolved DW and RMPS results are shown in \Fig{fig:terms}. We see that for both methods the cross-sections are larger for the \singS\ than the \tripS\ across the entire energy range. Furthermore, the DW results are approximately 15\% higher than the fitted RMPS results. The data fits were made using the formula of \citeasnoun{younger81pra}:
\begin{eqnarray}
\label{eq:fits}
\sigma_\textrm{ionization} = \frac{1}{IE} &\big(& A(1-1/u) + B (1-1/u)^2 \nonumber \\
 &&+ C\ln(u) + D\ln(u)/u \,\big)
\end{eqnarray}
where $I$ is the ionization energy, $E$ is the incident electron energy, and $u=E/I$. The coefficients $A$, $B$, $C$, and $D$ are determined from a least-squares fit to the calculated cross-section (fitting parameters are available from the authors upon request). Note that the coefficient $C$ can be independently determined from the photoionization cross-section, but we have left it as a free parameter. 

\begin{figure}[tb]
\centering
\includegraphics[width=0.75\textwidth]{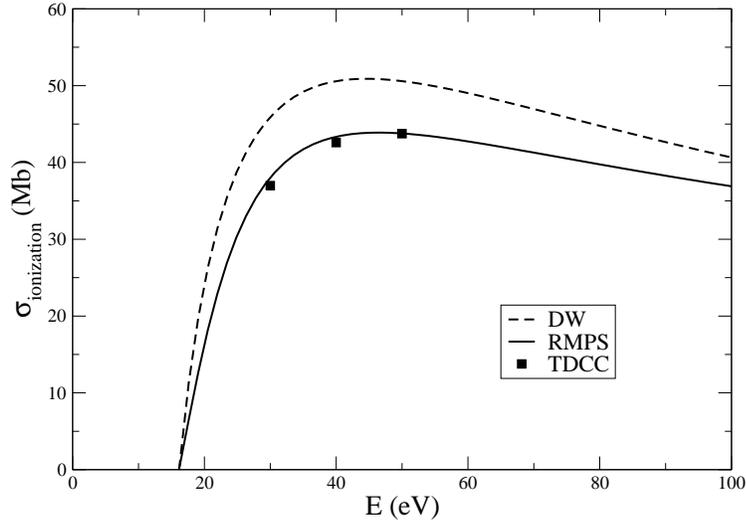}
\caption{\label{fig:config_av} Configuration-averaged electron-impact ionization cross-sections, $\sigma_\textrm{ionization}$, against incident electron energy, $E$. The distorted-wave calculation (dashed line) gives a cross-section around 15\% larger than the non-perturbative RMPS calculation (solid line). Three TDCC points are also shown (squares), and these are in good agreement with the RMPS calculation. \mbox{($1\,\textrm{Mb} = 10^{-18}\,\textrm{cm}^{2}$)}}
\end{figure}

While the DW and RMPS calculations are term resolved, and thus return both \singS\ and \tripS\ cross sections, the TDCC calculation is configuration averaged and only gives a single
cross-section for the $1s2s$ configuration. In order to compare the various theoretical calculations we show all of the configuration-averaged results in \Fig{fig:config_av}. The RMPS $^1\!S$ and $^3\!S$ results have been converted to a configuration-average cross section by averaging the fits of \eref{eq:fits}. The DW cross-section in this graph was calculated using the configuration-average approximation (see \Sec{ssec:dw}). We also took an average of the term resolved DW calculations, and it is in good agreement with the configuration-average result. The TDCC and RMPS cross sections are in excellent agreement with each other, while the DW result is around 15\% higher at the peak than the non-perturbative results.

It is interesting that the perturbative DW method produces cross-sections that are in good agreement with TDCC and RMPS for the ground state ($1s^2\ ^1S$) ionization \cite{pindzola00praA} and overestimates the metastable cross section. This is not unexpected for such a near neutral system, since DW doesn't include higher-order coupling between the outgoing electrons. This ``three body'' effect is more important when the  ionization limit is lower, as it is in the case of metastable \Liplus. This trend is similar to that found by \citeasnoun{griffin05jpb} for hydrogenic systems, including Li$^{2+}$. The discrepancy between the perturbative distorted-wave method and the non-perturbative RMPS and TDCC methods is therefore larger for ionization from excited states than for ground states.

One would also expect the effect to be more important when the effective ion charge (the charge that the escaping electrons see) is smaller. In fact, this trend can be seen when one compares the metastable He ionization cross-sections of \citeasnoun{colgan02pra} with the He-like Li results of this paper. As mentioned in the introduction, the DW and perturbative methods differ by a factor of two for metastable He. Furthermore, this trend can be observed by comparing ionization from H to ionization from Li$^{2+}$ \cite{griffin05jpb}.

\section{Conclusion}

We present results of three different calculations of the electron-impact cross section of metastable \singS\ and \tripS~\Liplus: distorted-wave, $R$-matrix with pseudostates, and time-dependent close-coupling. We find that the non-perturbative methods, $R$-matrix and time-dependent close-coupling, are in excellent agreement. The perturbative distorted-wave method, however, gives cross-sections that are significantly larger than those given by the non-perturbative methods. The non-perturbative \mbox{$R$-matrix} and close-coupling calculations provide a benchmark for future experimental determination of absolute ionization cross-sections from metastable states along the He isoelectronic sequence.

The fact that the DW and TDCC methods are actually in agreement for ionization cross-section from the \Liplus\ ground state \cite{pindzola00praA} has previously been taken as a sign that DW is fairly dependable for this species. However higher-order coupling between the outgoing electrons is not taken into account in the DW calculations. Although these correlations become smaller as the nuclear charge $Z$ increases, they become larger for excited-state ionization. Thus the non-perturbative methods are more appropriate for excited-state ionization. All existing electron-impact ionization data for \Liplus\ excited states comes from distorted-wave calculations \cite{loch06adndt}. In light of the results presented in this paper, however, this data will need significant revision.

\ack

This work was supported in part by grants from the US Department of Energy. Computational work was carried out at the National Energy Research Scientific Computing Center in Oakland, California, and at the National Center for Computational Sciences in Oak Ridge, Tennessee.

\section*{References}

\bibliographystyle{jphysicsB}
\bibliography{references}

\end{document}